\documentclass[preprint,onecolumn,groupedaddress,floatfix,showpacs,superscriptaddress]{revtex4}

\usepackage{amsmath}
\usepackage{amssymb}
\usepackage{natbib}
\usepackage{color}
\usepackage{graphicx}
\usepackage{hyperref}

\begin{document}
\bibliographystyle{apsrev}

\title{Equilibrium state of a cylindrical particle with flat ends in nematic liquid crystals}

\author{S. Masoomeh Hashemi}
\email{mhashemi@physics.sharif.edu}
\affiliation{Department of Physics, Sharif University of Technology, P.O. Box 11155-9161, Tehran, Iran}

\author{Mohammad Reza Ejtehadi}
\email{ejtehadi@sharif.edu}
\affiliation{Department of Physics, Sharif University of Technology, P.O. Box 11155-9161, Tehran, Iran}

\begin{abstract}
A continuum theory is employed to numerically study the equilibrium orientation and defect structures of a circular cylindrical particle with flat ends under a homeotropic anchoring condition in a uniform nematic medium. Different aspect ratios of this colloidal geometry from thin discotic to long rod-like shapes and several colloidal length scales ranging from mesoscale to nanoscale are investigated. We show that the equilibrium state of this colloidal geometry is sensitive to the two geometrical parameters: aspect ratio and length scale of the particle. For a large enough mesoscopic particle, there is a specific asymptotic equilibrium angle associated to each aspect ratio. Upon reducing the particle size to nanoscale, the equilibrium angle follows a descending or ascending trend in such a way that the equilibrium angle of a particle with the aspect ratio bigger than 1:1 (a discotic particle) goes to a parallel alignment with respect to the far field nematic, whereas the equilibrium angle for a particle with the aspect ratio 1:1 and smaller (a rod-like particle) tends toward a perpendicular alignment to the uniform nematic direction. The discrepancy between the equilibrium angles of the mesoscopic and nanoscopic particles originates from the significant differences between their defect structures. The possible defect structures related to mesoscopic and nanoscopic colloidal particles of this geometry are also introduced.
\end{abstract}

\date{\today}
\pacs{ 61.30.Cz, 82.70.Dd, 61.30.Jf, 61.30.Gd}
\maketitle

\section{Introduction}
Since the recognition of a novel class of long range intercolloidal interactions mediated by the elasticity of the nematic liquid crystal host~\cite{PoulinScience1997}, extensive research has been devoted to colloidal particles in uniform nematic media. When colloidal particles with definite boundary conditions are dispersed in liquid crystal media, localized regions with very low molecular order (topological defect cores) are formed in some specific places near the colloidal surfaces. These topological defect cores are accompanied by long-range slow-varying elastic interactions which are highly anisotropic even for the very symmetric geometries of spherical particles. For an individual spherical particle with a homeotropic anchoring condition immersed in a uniform nematic, two types of topological defects are in general known: a dipolar defect (a hyperbolic hedgehog point defect), and a quadrupolar defect (a disclination ring with winding number $-1/2$ called the Saturn ring).

Introducing anisotropy to the shape of colloidal particles dictates more complexity to the defect structures and interactions of a liquid crystal colloidal system. Defect structures specifically depend on the geometry of the particles. Therefore, by carefully choosing the geometrical shapes of the particles one can be able to design defect structures with predetermined architecture in a liquid crystal colloidal system possibly resulting in exclusive properties for technological applications.  Micro- and nano-scale colloidal particles with a variety of geometrical shapes are fabricated and studied in nematic liquid crystals~\cite{HaysMCLC1976, AndrienkoMDMC2002, TkalecMusevicRodLike2008, LapointeScience2009, MartinezMonopole2012,Microbullet2013, LapointeStar2013, TopologicalColloids2013}, but some crucial questions remain to be answered. The most important questions are about the equilibrium orientation of an anisotropic particle and the defect structures formed in the system. Spherocylindrical and elongated elliptical particles are well studied in a uniform nematic~\cite{AndrienkoMDMC2002,HungeAnisotropic2006,HungQuadrupolar2009,SmalyukhPRL2012,Tasinkevych2014}. In case of a homeotropic anchoring condition, both of these elongated colloidal particles orient with their symmetry axes perpendicular to the far field nematic. The perpendicular orientation has been confirmed using an experiment~\cite{SmalyukhPRL2012}, a combination of Monte Carlo and molecular dynamics simulations~\cite{AndrienkoMDMC2002} and a continuum theory~\cite{HungeAnisotropic2006,HungQuadrupolar2009,Tasinkevych2014}. In the case of very thin discotic particles with homeotropic anchoring a parallel alignment has been reported~\cite{HaysMCLC1976,HighAspectRatioSmalyukh2011, RovnerDisk2012}. However, for more complex colloidal shapes the stable orientation is not always a perpendicular or parallel alignment. While an elongated triangular nano-prism aligns perpendicular to the far field nematic, a cube with a homeotropic boundary condition has several energetically equivalent stable orientations~\cite{HungFaceted2009}. 

A number of parameters are thought to have controlling roles in the equilibrium state of cylindrical colloidal particles with a homeotropic anchoring condition in a uniform nematic. The free energy expressions for perpendicular and parallel alignments of an infinitely long cylindrical particle corresponding to rigid~\cite{BrochardGennes1970}, weak~\cite{BurylovPLA1990} and arbitrary~\cite{BurylovPRE1994, BurylovPRE2013} anchoring strengths have been analytically carried out using a continuum theory. It has been shown that a dimensionless parameter, which is constructed from the particle radius and liquid crystal elastic and anchoring constants, determines that the equilibrium state is either a parallel or a perpendicular orientation~\cite{BurylovPRE1994}. For finite-sized cylindrical colloidal particles the shape of the two ends has a pronounced effect in their equilibrium states~\cite{TkalecMusevicRodLike2008}. Although there are a number of studies discussing the equilibrium orientation and defect structures of cylindrical particles with flat ends ~\cite{TkalecMusevicRodLike2008,MatthiasCilyndricalParticle2009,HungFaceted2009,SenyukSmalyukhNanolett2012,RovnerDisk2012,BellerOrientation2015,Nikkhou2015}, there is no clear precise answer to the question considering various aspect ratios (ratio of diameter to height), length scales and anchoring strengths. Micron-sized rod-like solid particles with the typical aspect ratio 1:5 restricted to strong homeotropic anchoring conditions have been experimentally studied~\cite{TkalecMusevicRodLike2008}. The stability of the reported dipolar and quadrupolar defect structures depends on the orientation of the particle's symmetry axis with respect to the far field uniform nematic. The dipolar defect structures are stable in a small angular interval around the parallel orientation, while the quadrupolar defect structures are stable within a wider angular interval around the perpendicular alignment. 

Analytical description of liquid crystal colloidal systems has been in development along with experimental, numerical and simulation studies. An analytical description of arbitrary colloidal shapes in liquid crystals has been developed mostly in analogy with the common methods and expressions of electrostatics. Different types of elastic multipoles, specifically dipoles and quadrupoles, are characterized based on the symmetries in the director field around the colloidal surface~\cite{pergamenshchik2010,Pergamenshchik2011}. Using multipole expansion methods that are valid in regions without nonlinearities of the director field, pair and collective colloidal interactions are also discussed~\cite{Lev2001,Lev2002,Lev2012,pergamenshchik2010,Pergamenshchik2011}. However in close vicinity of the colloidal surface with nonlinearities of the director field no analytical description is achieved yet. Numerical minimization methods can be used in order to study the director field near a colloidal surface.

In this paper, using a finite element method for numerical minimization of the Lundau-de Gennes free energy, we present an almost detailed study on a circular cylindrical particle with flat ends that is immersed in a uniform nematic medium. A homeotropic boundary condition is imposed on the colloidal surface. We investigate the dependency of the equilibrium angle on the aspect ratio and length scale of the particle in the range of intermediate and strong anchoring strengths. Therefore, we study several mesoscopic and nanoscopic colloidal length scales and different aspect ratios ranging from thin discotic to long rod-like shapes. Some of the results are also compared with some previous experimental studies~\cite{TkalecMusevicRodLike2008, MatthiasCilyndricalParticle2009}. The defect structures related to different length scales are also introduced.

\section{Model}\label{model}

\subsection{Geometry}\label{Geometry}
\begin{figure}[!ht]
\begin{tabular}{c}
\includegraphics[width=7.0cm, angle=0]{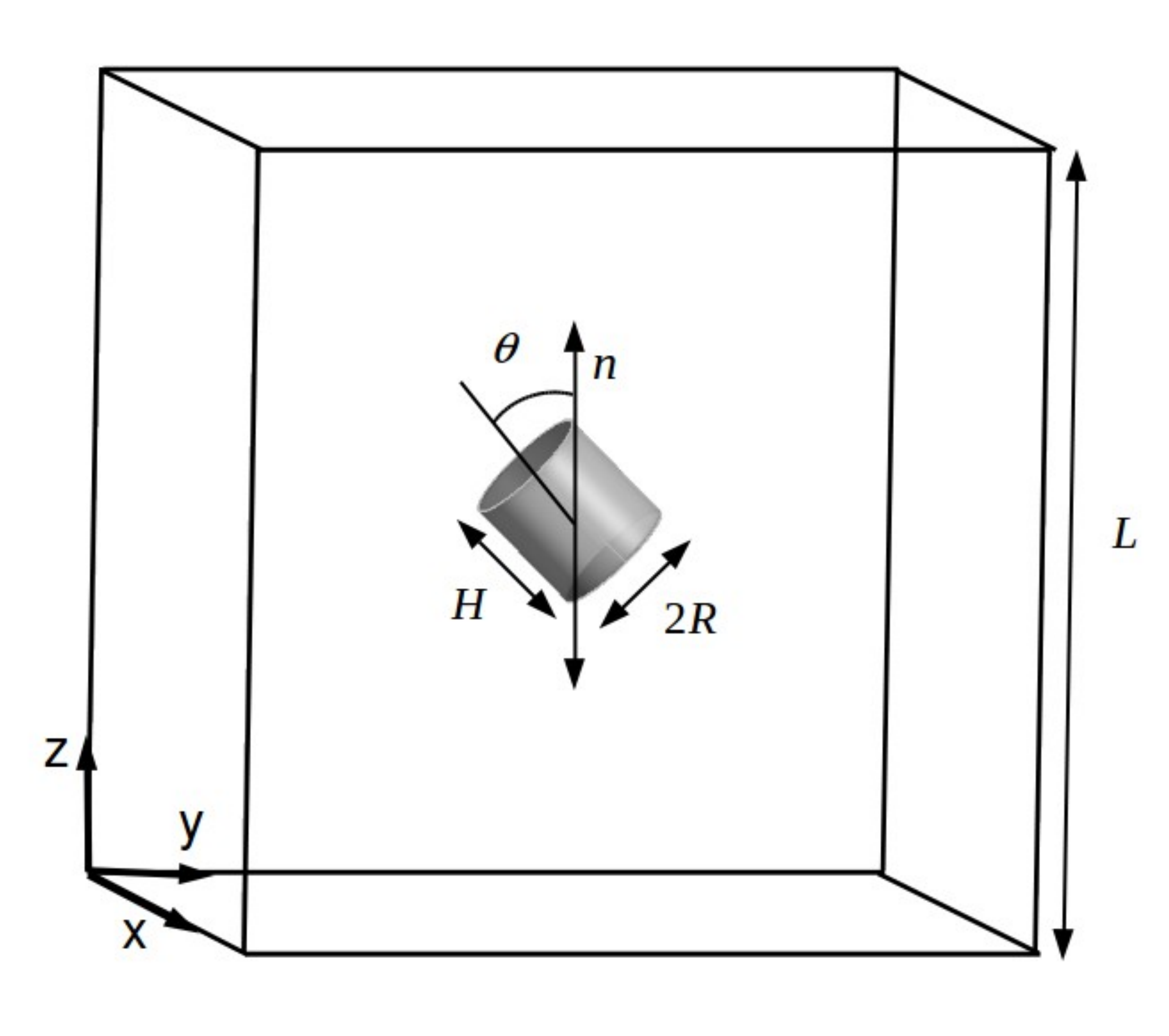}
\end{tabular}
\caption{Schematic illustration of the studied system.} \label{schem}
\end{figure}

We consider a cylindrical particle with a circular cross section and flat ends in a uniform field of nematic liquid crystal. The uniform nematic medium is modeled using a cubic box with a fixed anchoring condition in the $z$ direction at all the six walls of the cube. A schematic illustration of the system is shown in Fig.~\ref{schem}. 
To prohibit finite simulation box effects, the edge lengths of the cubic box (L) are taken to be 9 times bigger than the longest length scale of the cylindrical particle and the center of the particle is considered at the center of the simulation box. By this procedure the colloidal particle has an azimuthal symmetry at the center of the box (see Appendix A for more explanations). Therefore, the angular dependency of the free energy is only on the angle between the symmetry axis of the cylinder and the far field nematic direction ($\theta$). Therefore it is enough to consider the symmetry axis of the cylinder in the $y-z$ plane and let it be free to rotate around the $x$ axis. It is also sufficient to limit the investigation to the angular interval of $\theta$ from 0 to 90 degrees.

\subsection{Free Energy Expression}\label{free energy}

The spatially varying orientations of nematic liquid crystal molecules in a colloidal system can be described using the symmetric traceless tensor order parameter $Q_{ij}$ 
\begin{equation}
\begin{split}
Q_{ij}=S(n_in_j-\frac{\delta_{ij}}{3})
\end{split}
\label{scalarOrder}
\end{equation}
where the director $\vec{n}$ represents the local average orientations of liquid crystal molecules and the scalar order parameter $S$ gives a scalar measure of the local degree of molecular orientational order along the director. To avoid dealing with singularities of the defect cores, instead of working with the director $\vec{n}$ in numerical minimization, we use the general form of the tensor order parameter with six components $Q_{xx}, Q_{yy}, Q_{zz}, Q_{xy}, Q_{xz}$ and $Q_{yz}$, in which $x, y$ and $z$ represent the Cartesian coordinates. The largest eigenvalue of the tensor order parameter gives the scalar order parameter and the principal eigenvector corresponding to this eigenvalue gives the director of the nematic field.

To obtain the equilibrium nematic field, we minimize the free energy of the system which is written, based on the Lundau-de Gennes model~\cite{deGennes1995}, in powers of the tensor order parameter and its derivatives. A summation over the Lundau-de Gennes free energy for the isotropic-nematic phase transition, $f_{IN}$, the elastic free energy for the uniaxial nematic~\cite{KraljZumer1992}, $f_{el}$, and the surface energy for imposing a specific anchoring direction to the directors at the colloidal surfaces~\cite{Nobili1992}, $f_{s}$, yields a general expression for the free energy,
\begin{equation}
\begin{split}
F &=\int_{\mathrm{bulk}} \mathrm{d}V \ (f_{\mathrm{IN}}+f_{\mathrm{el}}) + \int_{\mathrm{col.\ sur.}} \mathrm{d}A \ f_{\mathrm{s}},
\end{split}
\label{generalEnergy}
\end{equation}
The Landau-de Gennes free energy for the isotropic-nematic phase transition is written in powers of the tensor order parameter as
\begin{equation}
\begin{split}
f_{\mathrm{IN}} &=\frac{A(T)}{2}Q_{ij}Q_{ji}-\frac{B}{3}Q_{ij}Q_{jk}Q_{ki}+\frac{C}{4}\left(Q_{ij}Q_{ji}\right)^2,
\end{split}
\label{INEnergy}
\end{equation}
where $A, B$ and $C$ are material-dependent parameters and only $A(T)$ is taken to be temperature dependent. 

The elastic energy for uniaxial nematic is constructed from the tensor order parameter derivatives in the form
\begin{equation}
\begin{split}
f_{\mathrm{el}} &= \frac{L_1}{2}\partial_{k}Q_{ij}\partial_{k}Q_{ij}+\frac{L_2}{2}\partial_{k}Q_{ik}\partial_{j}Q_{ij}\\ &+\frac{L_3}{2}Q_{ij}\partial_{i}Q_{kl}\partial_{j}Q_{kl}+\frac{L_4}{2}\partial_{j}Q_{ik}\partial_{k}Q_{ij},
\end{split}
\label{elasticEnergy}
\end{equation}
in which the elastic constants $L_1, L_2, L_3$ and $L_4$ are related to the Frank-Oseen elastic constants $K_1, K_2, K_3$ and $K_{24}$ (respectively denoting splay, twist, bend and saddle-splay elastic constants) by
\begin{equation}
\begin{split}
L_1 &= \frac{3K_{2}+K_{3}-K_{1}}{6S^2},\\
L_2 &= \frac{K_{1}-2K_{24}}{S^2},\\
L_3 &= \frac{K_{3}-K_{1}}{2S^3},\\
L_4 &= \frac{-K_{2}+2K_{24}}{S^2},
\end{split}
\end{equation}
We adopt one elastic constant approximation by taking $K_{1}=K_{2}=K_{3}=2K_{24}$ that sets $L_2, L_3$ and $L_4$ equal to zero.

The surface energy for a preferential anchoring of the director on the colloidal surfaces is
\begin{equation}
\begin{split}
f_{\mathrm{s}} &=\frac{W}{2} \left(Q_{ij}-Q^{\mathrm{s}}_{ij}\right)\left(Q_{ji}-Q^{\mathrm{s}}_{ji}\right)
\end{split}
\label{surfaceEnergy}
\end{equation}
where $W$ is the anchoring constant and $Q^{\mathrm{s}}$ is referred to the preferred orientational anchoring of the molecules at the colloidal surfaces. We assume a homeotropic anchoring with $Q^{\mathrm{s}}_{ij}=S(\nu_i\nu_j-\delta_{ij} / 3)$ in which $S$ is the bulk scalar order parameter and $\vec{\nu}$ stands for the normal vector to the colloidal surfaces.  

A combination of rescaling parameters and changing variables is a useful technique to modify the equations to the ones with more appropriate units and fewer material dependent parameters which are favorable to numerical calculations.  We rescale the scalar order parameter as $S = S_{eq} S_{q}$ and subsequently the tensor order parameter as $Q = S_{eq} q$, where $S_{eq}=(\frac{2}{3})^{3/2} B/C$ is the dimensionless rescaling parameter and $q=S_{q} (n_in_j-\delta_{ij}/3)$ is the rescaled tensor order parameter. The dimensionless free energy of the system becomes
\begin{equation}
\begin{split}
\frac{F}{f_{0}R^{3}} &=\int \frac{\mathrm{d}V}{R^3} \bigg(\frac{\tau }{2}q_{ij}q_{ji}-\frac{\sqrt{6}}{4}q_{ij}q_{jk}q_{ki}+\frac{1}{4}\left(q_{ij}q_{ji}\right)^2+\frac{1}{2}(\frac{\xi}{R})^{2}\hat{\partial}_{k}q_{ij}\hat{\partial}_{k}q_{ij} \bigg)\\ & + \int \frac{\mathrm{d}A}{R^2} \bigg(\frac{1}{2} (\frac{\eta}{R}) \left(q_{ij}-q^{\mathrm{s}}_{ij}\right)\left(q_{ji}-q^{\mathrm{s}}_{ji}\right) \bigg),
\end{split}
\label{dimensionlessEnergy}
\end{equation}
where $f_{0}=C S_{eq}^{4}$ and $\hat{\partial}=R\partial$. The dimensionless free energy involves only three 
material dependent parameters: the dimensionless temperature $\tau=A/C S_{eq}^{2}$, the dimensionless elastic constant $\xi/R=\sqrt{L_1/C R^2 S_{eq}^{2}}$ and the dimensionless anchoring constant $\eta/R=W/C R S_{eq}^{2}$. A first order isotropic-nematic phase transition takes place at $\tau=1/8$ and the isotropic phase is unstable for $\tau<0$. We take the equilibrium scalar order parameter of the uniform bulk nematic equal to $S_{eq}$, so we have $\tau=(3\sqrt{6}-8)/12$. In this study we adopt the same parameters and rescaling method as Ref.~\cite{FukudaPhysRevE2004}, so we have $S_{eq}=0.653$ and we always take the elastic constant to be $L_1=25\times 10^{-12} \rm{J/m}$. There are two accepted conventions for estimating the correlation length. One definition is taken from the rescaled elastic constant in the above mentioned procedure, $\xi$, which is the same definition as introduced in Ref.~\cite{deGennes1995}. The other definition comes from the correlation function definition~\cite{chandrasekhar1992liquid}. Both definitions give similar estimates for the correlation length. Based on these two definitions the correlation length in this study is $16.5\ \rm{nm}$ or $14\ \rm{nm}$ ~\cite{correlation}. To estimate the limits of weak, intermediate and strong anchoring strengths, the dimensionless parameter
\begin{equation}
w=WR/L_1
\label{dimensionlessAnchoring}
\end{equation}
provides a criterion for comparing the relative contribution of the surface energy to the elastic energy~\cite{deGennes1995}. $R$ is the particle's characteristic length which is taken, in this study, to be equal to the particle radius. Note that $w\ll 1$ implies a weak anchoring limit whereas $w\gg 1$ describes a strong anchoring condition.

\subsection{Numerical Minimization of the Free Energy}\label{model}

To numerically minimize the total free energy of the model system, we implement a finite element method in which the nematic bulk of the system is discretized into tetrahedral elements. To do this, we have used the three-dimensional (3D) finite element mesh generator~\emph{Gmsh}~\cite{Gmsh}. To reduce the minimization time simultaneously with having sufficient accuracy near the colloidal surface, a much finer mesh resolution is applied on the colloidal surface than on the cubic box of the model. Starting from given initial conditions, the numerical minimization is carried out using a conjugate gradient (CG) method~\cite{Press.92}. The angular dependency of the free energy is scanned in the angular interval $0\leq \theta\leq 90$ in steps of 5 degrees. 
The minimization process continues until the free energy difference of the subsequent director configurations drop below about $0.0001\%$. We have studied cylindrical colloidal particles with rounded edges to achieve more efficient numerical minimization which is the same approach as that adopted in Ref.~\cite{HungFaceted2009}. The radius of the rounded edges is assumed to be much smaller than the height (H) and radius (R) of the colloidal particles and the liquid crystal correlation length.

\section{Results and Discussion}

We use the term \textit{aspect ratio} to refer to the ratio of the diameter to the height of the cylindrical particle ($2R/H$). 
We will investigate the equilibrium orientation concomitantly with the defect structures for the colloidal dimensions from tens to a few hundreds of nanometers. Several colloidal shapes with the mentioned geometry but different aspect ratios are considered. Hence our investigation includes various colloidal shapes from thin discotic to long rod-like cylindrical particles.

The general topography of the director field for this colloidal geometry consists of a disclination loop with winding number $-1/2$ that is tightly folded/wrapped around the colloidal surface. The disclination loop may experience an abrupt bending or piecewise widening that is strongly dependent on some particle-dependent parameters such as spatial orientation and length scale and some material-dependent parameters like anchoring and elastic constants. 

\begin{figure*}[!ht]
\begin{tabular}{cccccccc}
\includegraphics[width=14.0cm, angle=0]{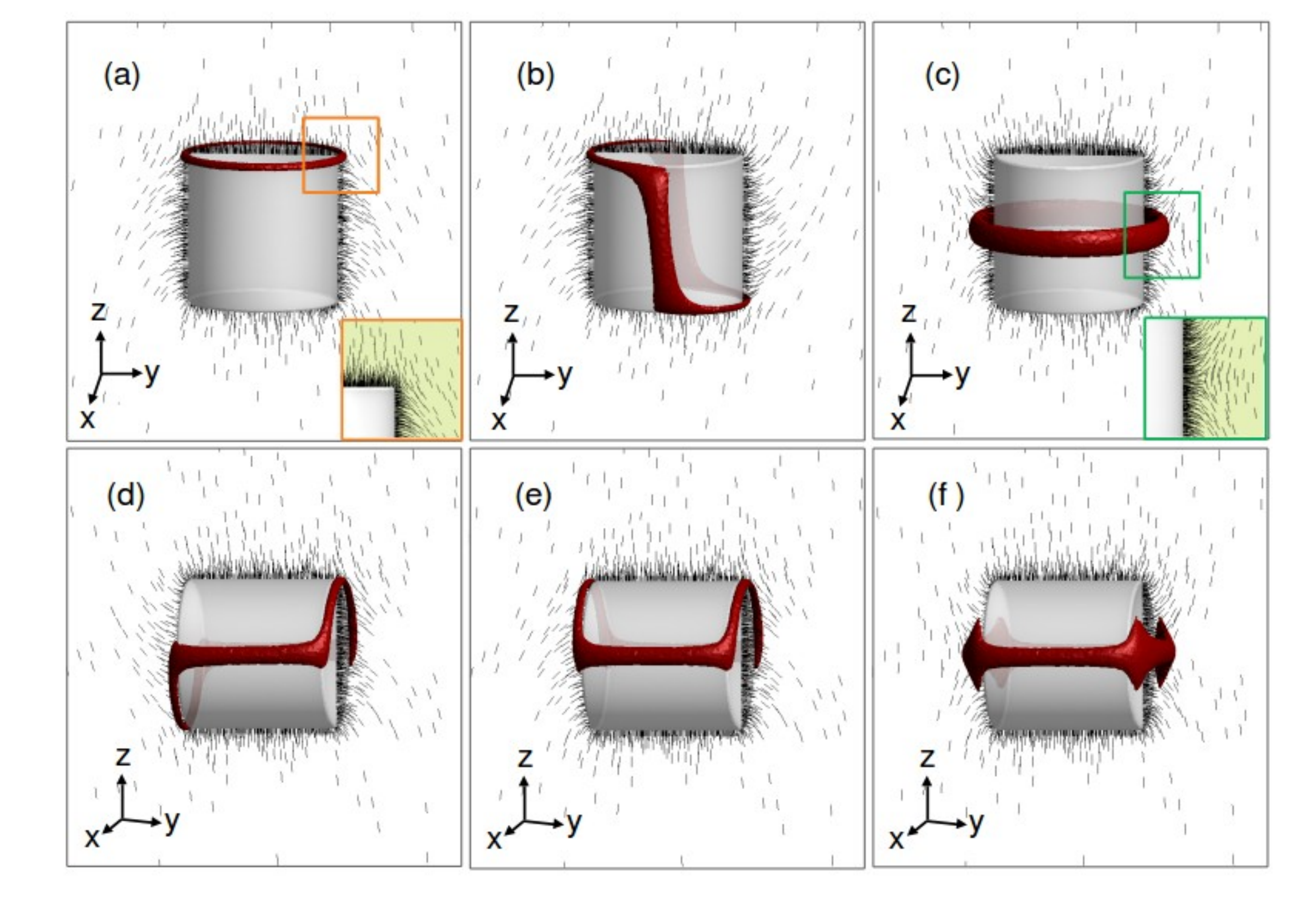}
\end{tabular}
\caption{(Color online) Defect structures of a mesoscopic cylindrical particle with $R=160 \rm{nm}$ and aspect ratio 1:1 for $W= 10^{-2} \rm{J/m^2}$. Black dashed lines are a 2D view of the director field in the $y-z$ plane. Topological defect cores are visualized by the isosurface $S_q=0.5$ of the rescaled bulk scalar order parameter. (a-c) Defect structures for $\theta=0$ in descending order of stability: (a) top-ring, (b) chair-like, and (c) mid-ring-0. (d-f) Defect structures for $\theta=90$ in descending order of stability: (a) chair-like, (b) boat-like, and (c) mid-ring-90. The insets of (a) and (c) illustrate the director field across the defect cores of the corresponding defect lines.} \label{defectGeneral}
\end{figure*}

The several types of defect structures we have observed for a mesoscopic colloidal particle with strong homeotropic anchoring strength ($w=64$), are depicted in Fig.~\ref{defectGeneral}.
The relative stability of these defect structures is highly dependent on the spatial orientation of the particle and can be generally indicated for all the range of aspect ratios we have studied here. For parallel alignment of the symmetry axis with respect to the far field nematic, we have observed the three defect structures of Figs.~\ref{defectGeneral}(a-c) among which the top-ring is the most stable structure (Fig.~\ref{defectGeneral}(a)). A hyperbolic point defect structure has been neither experimentally nor numerically reported for this colloidal geometry. The reported dipolar structure is a flat thin ring located at the top or bottom edge of the cylindrical particle~\cite{TkalecMusevicRodLike2008, MatthiasCilyndricalParticle2009} (Fig.~\ref{defectGeneral}(a)) which is called the top-ring structure here. The three structures for the parallel alignment are depicted in descending order of stability from the most to the least stable structures in Fig.~\ref{defectGeneral}(a-c). Thus the chair-like structure in Fig.~\ref{defectGeneral}(b) is the next stable structure with free energy higher than that of the top-ring by about $3.8\%$ and the mid-ring-0 in Fig.~\ref{defectGeneral}(c) has the highest energy cost with free energy higher than that of the top-ring by about $16\%$. For the perpendicular alignment, we have observed the three structures of Fig.~\ref{defectGeneral}(d-f) which are depicted in descending order of stability. The free energy of the boat-like structure of Fig.~\ref{defectGeneral}(e) is about $1.1\%$ and the mid-ring-90 of Fig.~\ref{defectGeneral}(f) about $7.6\%$ higher than that of the chair-like structure of Fig.~\ref{defectGeneral}(d). We have seen that the same order of stability as in Figs.~\ref{defectGeneral}(a-c) and Figs.~\ref{defectGeneral}(d-f) is true for the defect structures of all the mesoscopic particles studied here. The free energy difference between these structures depends on the aspect ratio and length scale of the particle and the anchoring strength. For oblique orientations of a mesoscopic particle from 0 to about 10 degrees the top-ring structure is the most stable structure and from about 10 to 90 degrees the chair-like structure is the most stable structure. In such a liquid crystal colloidal system, besides the very high elastic energy cost of the disclination core, the sharp edges of the particle also induce strong elastic deformations into the director field. Therefore it can be seen from the possible defect structures corresponding to each spatial orientation that superimposition of the topological defect cores on the sharp edges of the particle reduces the elastic energy cost resulting in more stable structures.

\begin{figure}[!ht]
\begin{tabular}{cccc}
\includegraphics[width=9.0cm, angle=0]{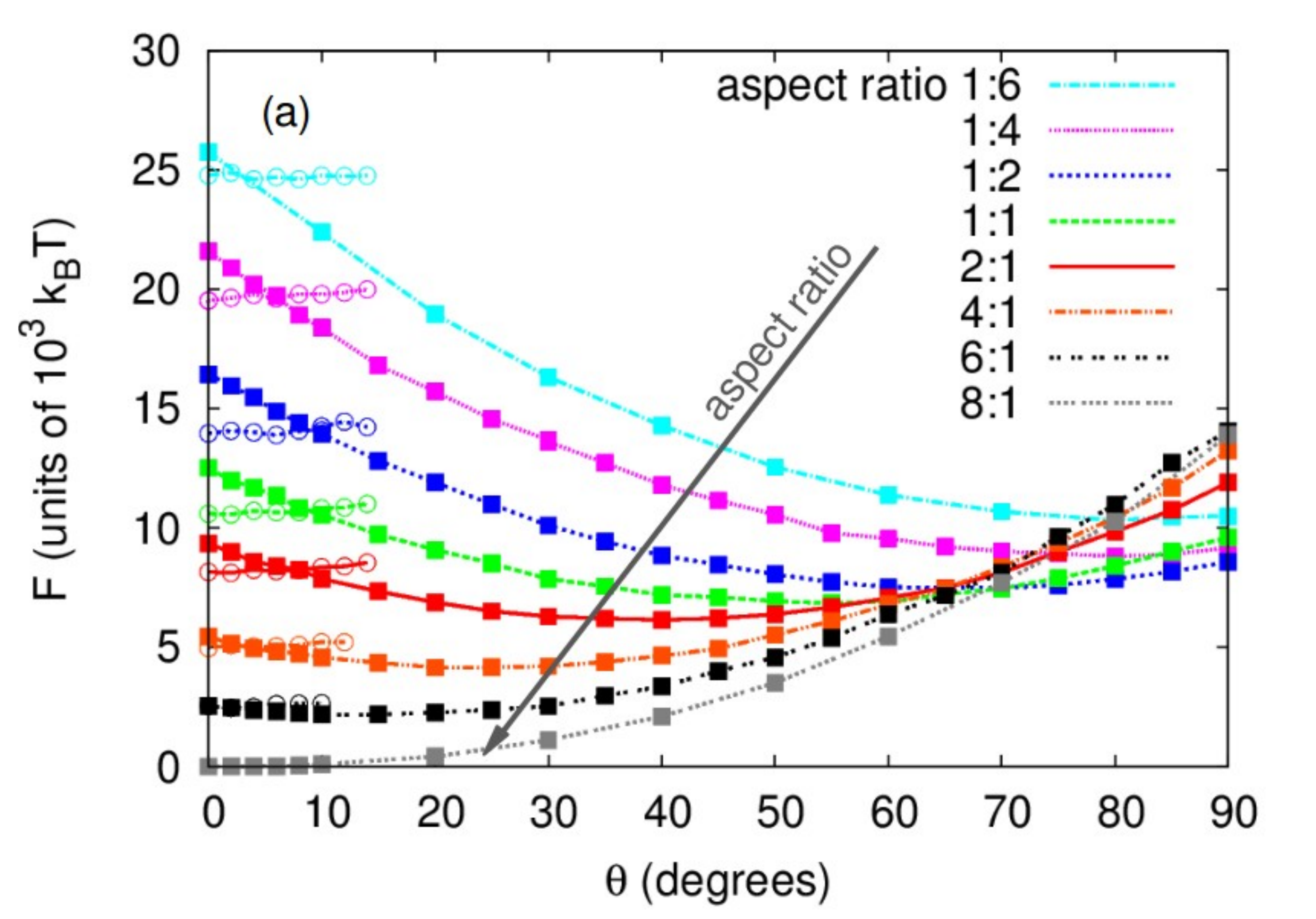}\\
\includegraphics[width=9.0cm, angle=0]{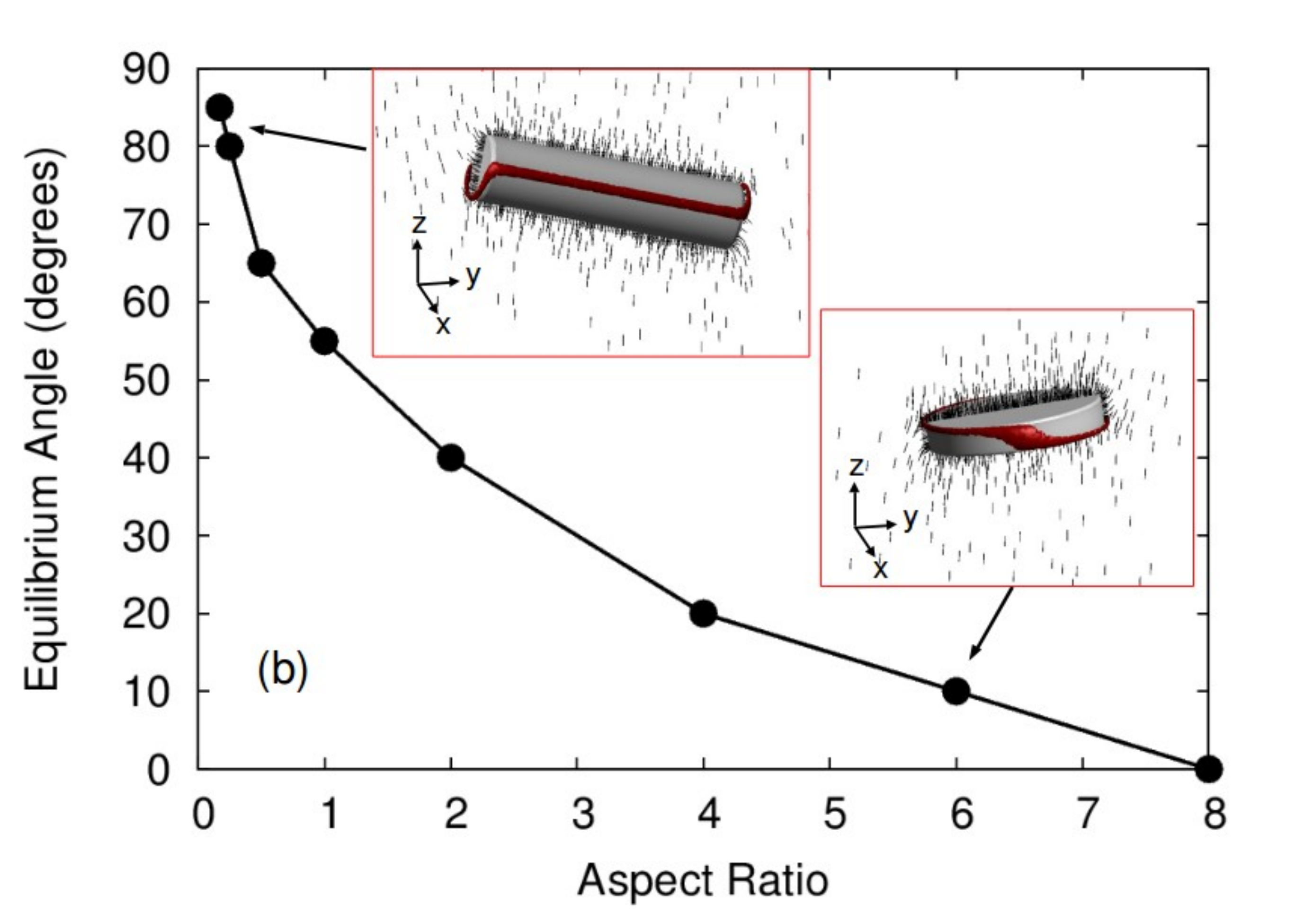}
\end{tabular}
\caption{(Color online) (a) Total free energy of the cylindrical particles with the aspect ratios from 8:1 to 1:6 as a function of $\theta$ for the homeotropic anchoring constant $W= 10^{-2} \rm{J/m^2}$. Solid squares are related to the chair-like structure and open circles represent the top-ring structure. All the particles have equal surface areas with a particle of the aspect ratio 1:2 and $R=160 \rm{nm}$. The angular resolution is 5 degrees for the chair-like structure and 2 degrees for the top-ring structure. (b) Equilibrium angle as a function of aspect ratio. The equilibrium defect structures for the aspect ratios 1:4 and 6:1 (as typical examples respectively for rod-like and discotic particles) are shown in the insets of (b) in which black dashed lines are 2D view of the director field in the $y-z$ plane and the topological defect cores are visualized by the isosurface $S_q=0.5$. } \label{equalSurface}
\end{figure}

If we restrict the colloidal particles of different aspect ratios to have equal surface areas, the free energy of all the systems including a thin discotic or a long rod-like particle become comparable to each other and all the energy graphs can be simultaneously shown in one single plot. This procedure helps us present some general results that we have obtained for the mesoscopic particles with the aspect ratios 8:1 to 1:6. The free energy graphs of these particles with respect to $\theta$ for strong homeotropic anchoring with the anchoring constant $W= 10^{-2} \rm{J/m^2}$ are illustrated in Fig.~\ref{equalSurface}(a).

It can be seen from the energy graphs of Fig.~\ref{equalSurface}(a) that, no matter which aspect ratio the particle has, the top-ring dipolar defect is the most stable structure within the angular interval from 0 to about 10 degrees. After a crossover around 10 degrees from the top-ring to the chair-like structure, the most stable structure from about 10 to 90 degrees is the chair-like quadrupolar structure. The same angular interval of stability for the top-ring structure has been experimentally observed for microrods of the typical aspect ratios 1:5 with strong homeotropic anchoring~\cite{TkalecMusevicRodLike2008}. This is true for all the aspect ratios from 8:1 to 1:6 studied here. This interval of stability for the top-ring defects is exclusive to the geometry we have considered whereas for an elongated elliptical particle the top-ring dipolar defect is stable within the wider angular interval from 0 up to about 40 degrees~\cite{Tasinkevych2014}. We have also observed the mid-ring-0 in a small angular interval around 0 degrees and the boat-like and the mid-ring-90 in a small angular interval around 90 degrees.

Another general point that can be deduced from the energy graphs of Fig.~\ref{equalSurface} is existence of only one global minimum corresponding to each aspect ratio in the interval from 0 to 90 degrees. It can be seen, Fig.~\ref{equalSurface}, that the equilibrium angle that is about 90 degrees for long rod-like colloidal particles gradually decreases upon increasing the aspect ratio and ultimately reaches about 0 degrees for thin discotic particles. The equilibrium angle for the aspect ratio 1:1 is $55\pm5$ degrees which increases to $85\pm5$ degrees for a long cylinder with the aspect ratio 1:6. The experimental study of Ref.~\cite{TkalecMusevicRodLike2008} on rod-like particles with the average aspect ratio 1:5 shows the same interval of stability for a quadrupolar structure. We see that the quadrupolar chair-like structure is the stable structure for aspect ratios of 1:1 and smaller in Fig.~\ref{equalSurface}(a). This is a characteristic of this colloidal geometry with flat ends, whereby there are different global minimums corresponding to different aspect ratios. In contrast, for example, in the case of a prolate elliptical particle with a homeotropic anchoring condition, the equilibrium alignment in a uniform nematic is a perpendicular alignment that is irrespective to the aspect ratio of the particle, the length scale of the particle and the anchoring strength~\cite{Tasinkevych2014}.

\begin{figure}[!ht]
\begin{tabular}{cccc}
\includegraphics[width=9.0cm, angle=0]{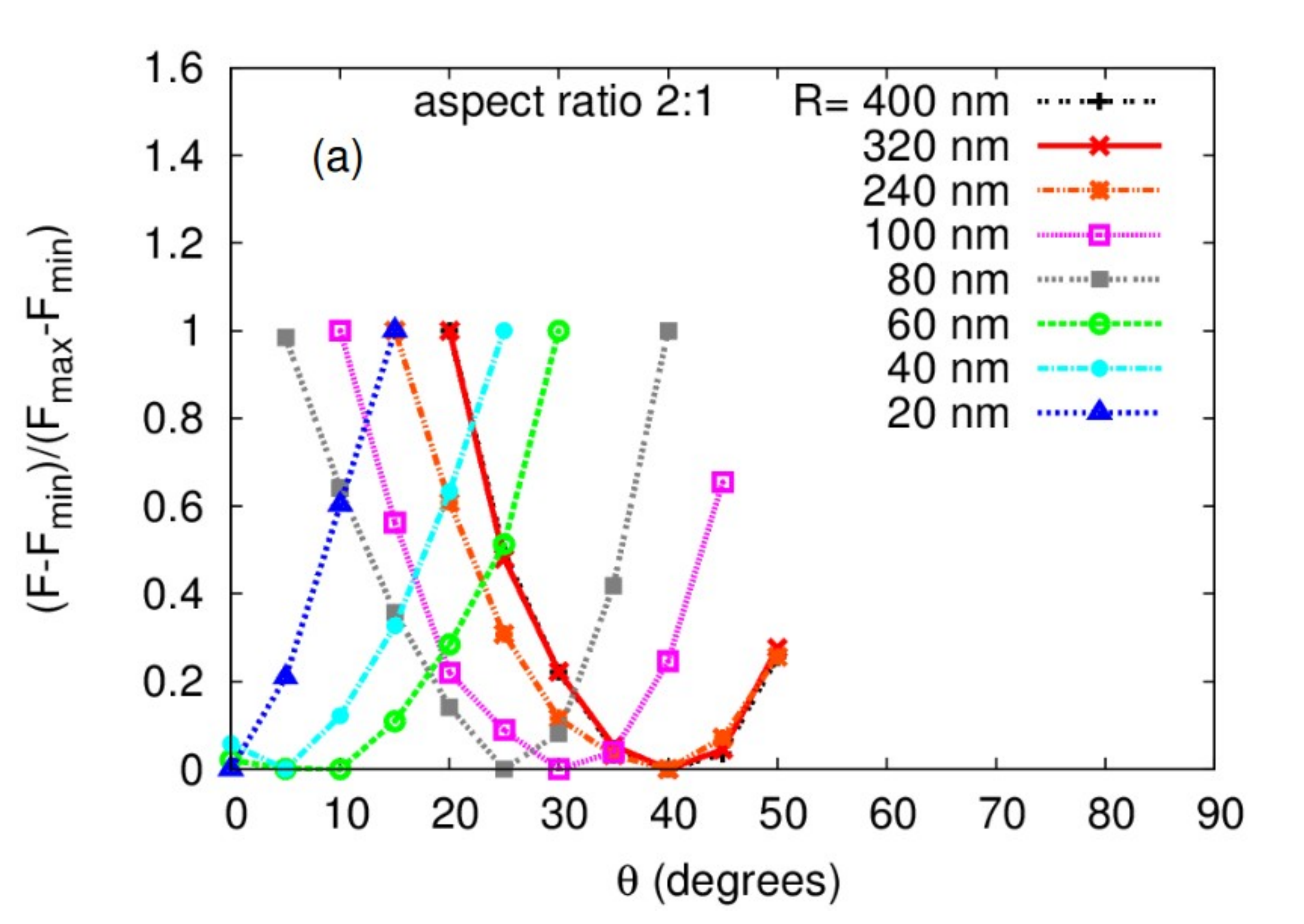}\\
\includegraphics[width=9.0cm, angle=0]{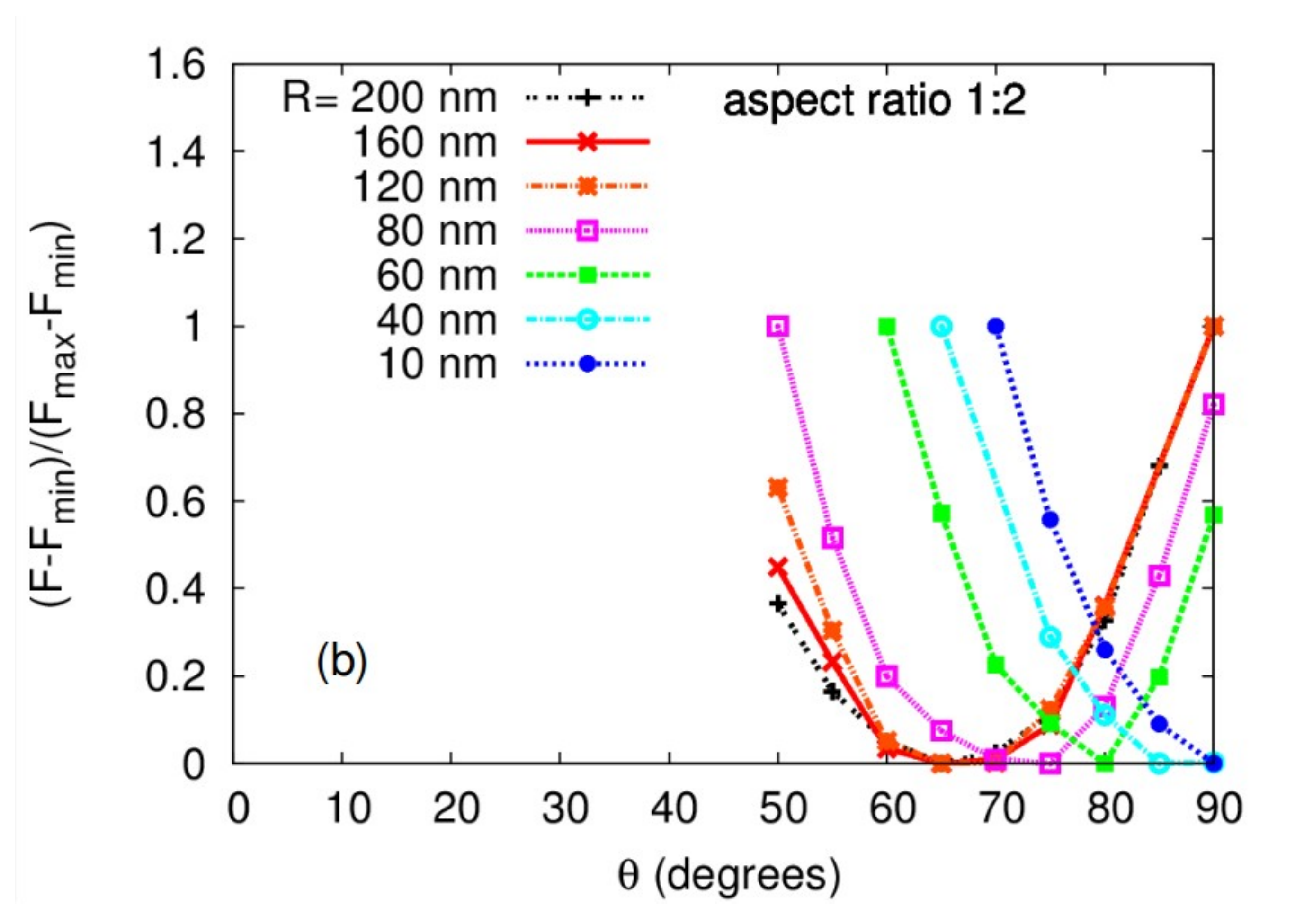}
\end{tabular}
\caption{(Color online) Normalized free energy as a function of $\theta$ when the particle radius changes from mesoscale to nanoscale considering $W=5\times 10^{-3} \rm{J/m^2}$ for (a) the aspect ratio 2:1 and (b) the aspect ratio 1:2.}
\label{sizeChange-part}
\end{figure}

The colloidal geometry studied here exhibits an exclusive equilibrium behavior according which the equilibrium angle depends on not only the aspect ratio but also the length scale of the particle. To clarify this point, we assume the material dependent parameters are constant, i.e. $L_1=25\times 10^{-12} \rm{J/m}$ and $W=5\times 10^{-3} \rm{J/m^2}$, while we change the particle radius from mesoscale to nanoscale. We firstly carry out calculations for the aspect ratios 2:1 and 1:2 (as typical examples respectively for the discotic and rod-like particles of this geometry). The energy graphs of Fig.~\ref{sizeChange-part} illustrate the energies around the global minimums corresponding to each particle radius. To make the comparison of the energies easier, we show the energy graphs just around their global minimums and normalize them. The normalization of the energies is carried out using the maximum ($F_{\rm{max}}$) and minimum ($F_{\rm{min}}$) value of each energy graph.

Figure ~\ref{sizeChange-part}(a), for the aspect ratio 2:1, shows that for all the mesoscopic colloidal particles with $R=400, 320$ and $240 \rm{nm}$ the equilibrium angle is $40\pm5$ degrees. By reducing the particle radius, the equilibrium angle gradually decreases and ultimately reaches $0\pm5$ degree for $R=20 \rm{nm}$. Figure ~\ref{sizeChange-part}(b) shows strikingly different behavior for the aspect ratio 1:2 upon reducing the particle size. In this case for all the mesoscopic colloidal particles with $R=200, 160$ and $120 \rm{nm}$ the equilibrium angle is $65\pm5$ degrees. In contrast to the previous case, for this aspect ratio by reducing the particle radius the equilibrium angle increases, so that for $R=10 \rm{nm}$ the equilibrium angle reaches $90\pm5$ degrees. Therefore the equilibrium angles associated to different aspect ratios respond differently to the changes in the particle size. 
 
\begin{figure}[!ht] 
\begin{tabular}{cc}
\includegraphics[width=9.0cm, angle=0]{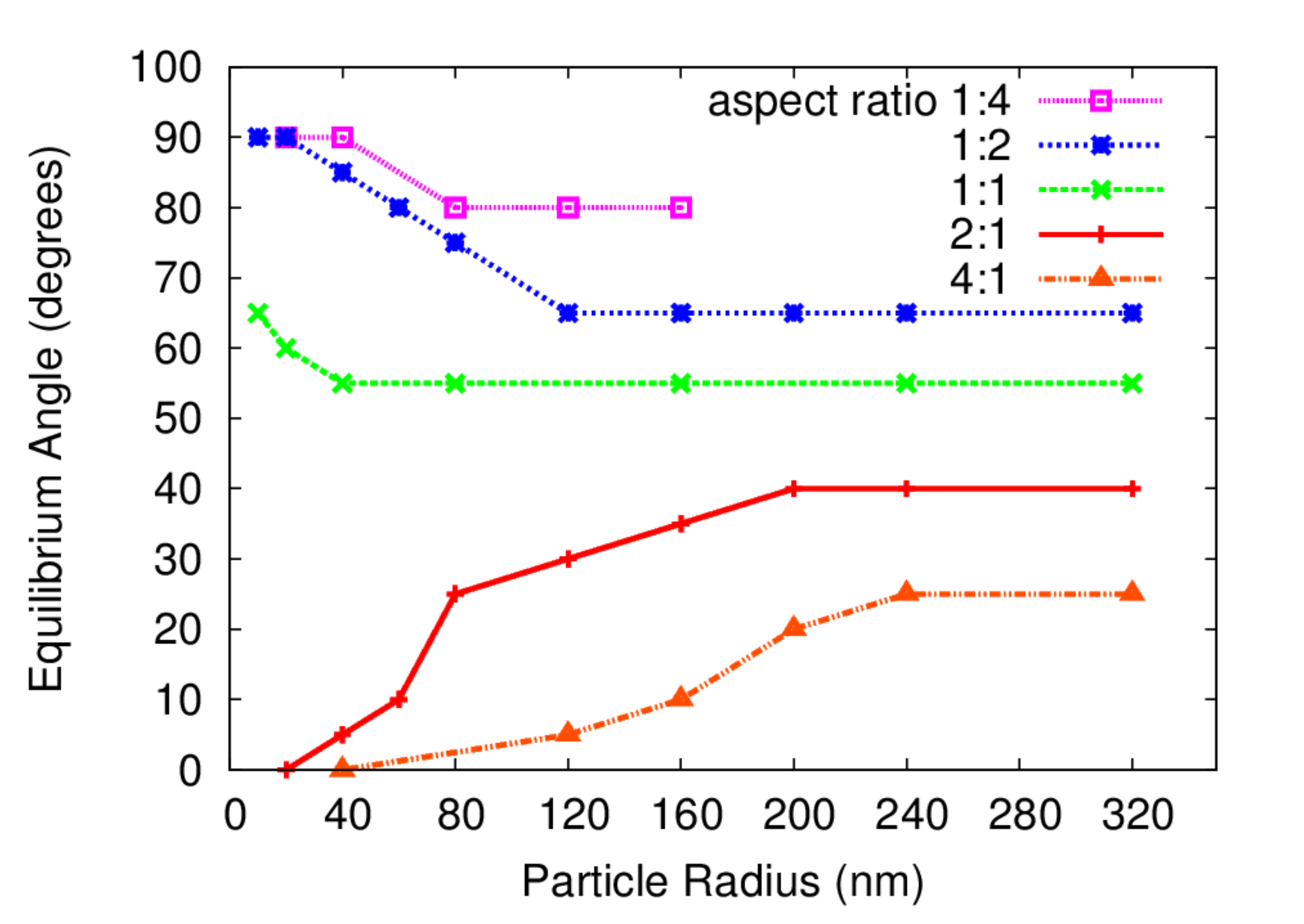}
\end{tabular}
\caption{(Color online) Equilibrium angle as a function of the particle radius for the anchoring constant $W=5\times 10^{-3} \rm{J/m^2}$. Small nanoparticles with $R=10 \rm{nm}$ ($w=2$) to mesoscopic particles with $R=320 \rm{nm}$ ($w=64$) are considered. The angular resolution for the equilibrium angles is 5 degrees.} 
\label{result}
\end{figure}

The graphs of Fig.~\ref{result} illustrate the equilibrium angles related to the aspect ratios 1:4, 1:2, 1:1, 2:1 and 4:1 as a function of the particle radius.
Figure ~\ref{result} shows that for all the mesoscopic colloidal particles of different aspect ratios, when the particle radius exceeds a specific value, the equilibrium angle reaches an asymptotic value. Note that there is a specific asymptotic equilibrium angle corresponding to each aspect ratio. As the results of Fig.~\ref{result} show, the equilibrium behavior of this colloidal geometry differs remarkably when the size of the colloidal particle reaches nanoscales. Based on the aspect ratio of the particle, there are two types of limiting behaviors. By reducing the particle radius to nanoscale, the equilibrium orientation for the particle with an aspect ratio larger than 1:1 (discotic particles) goes to a parallel alignment whereas the equilibrium orientation for the colloidal particle with the aspect ratio 1:1 or smaller (rod-like particles) tends toward a perpendicular alignment (Fig.~\ref{result}). 

\begin{figure*}[!ht]
\begin{tabular}{cccc}
\includegraphics[width=9.0cm, angle=0]{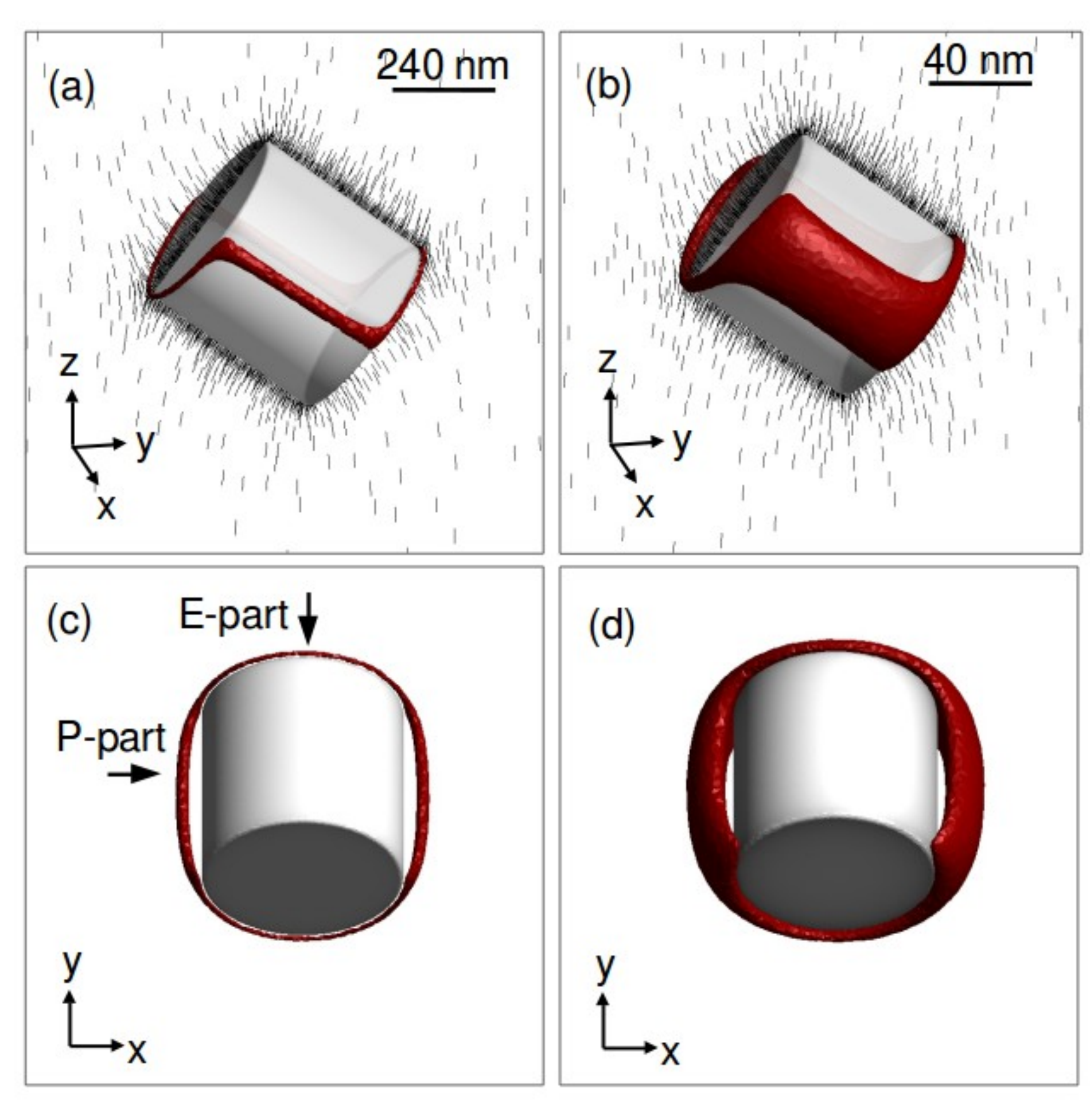}
\end{tabular}
\caption{(Color online) Equilibrium defect structures around colloidal particles with the aspect ratio 1:1 for (a) and (c) $R=240 \rm{nm}$ and (b) and (d) $R=40 \rm{nm}$ and $W=5\times 10^{-3} \rm{J/m^2}$. Black dashed lines in (a) and (b) are a 2D view of the director field in the $y-z$ plane. Topological defect cores are visualized by the isosurface $S_q=0.5$.} 
\label{nano-micro}
\end{figure*}

Given particle size changes from mesoscale to nanoscale have also pronounced effects on the defect structures. The equilibrium defect structures related to the aspect ratio 1:1 and $R=240 \rm{nm}$ and $R=40 \rm{nm}$ are depicted in Figs.~\ref{nano-micro}(a-d).
The disclination cores of both cases are well-localized in a small distance away from the colloidal surface. Figures ~\ref{nano-micro}(c) and (d) are another view of the disclination lines that better show the disclination core's distance from different parts of the colloidal surface. We call those parts of the chair-like structure that lie on the sharp edges the E-parts and those parts that are elongated parallel to the symmetry axis the P-parts. Figure ~\ref{nano-micro}(a) shows that the chair-like structure for $R=240 \rm{nm}$ is a thin line compared to the particle size. But the chair-like structure for $R=40 \rm{nm}$ consists of a much thicker line relative to the particle size for which the core size of its P-part is much wider than the E-part, Fig.~\ref{nano-micro}(b). Therefore, by reducing the particle size not only the defect core size relative to the particle size increases but also the thickness difference between the P-part and E-part of the defect line becomes more pronounced. This is true for the chair-like structure of any other aspect ratio. 

As we mentioned, we have observed three possible defect structures related to each of the parallel and perpendicular alignments of the mesoscopic colloidal particles (Fig.~\ref{defectGeneral}(a-f)), while by going to the limiting state for the nanoscopic colloidal particles (mentioned in analyzing the results of Fig.~\ref{result}) we have observed only one defect structure associated to each of these angles. 
\begin{figure*}[!ht]
\begin{tabular}{cccc}
\includegraphics[width=14.0cm, angle=0]{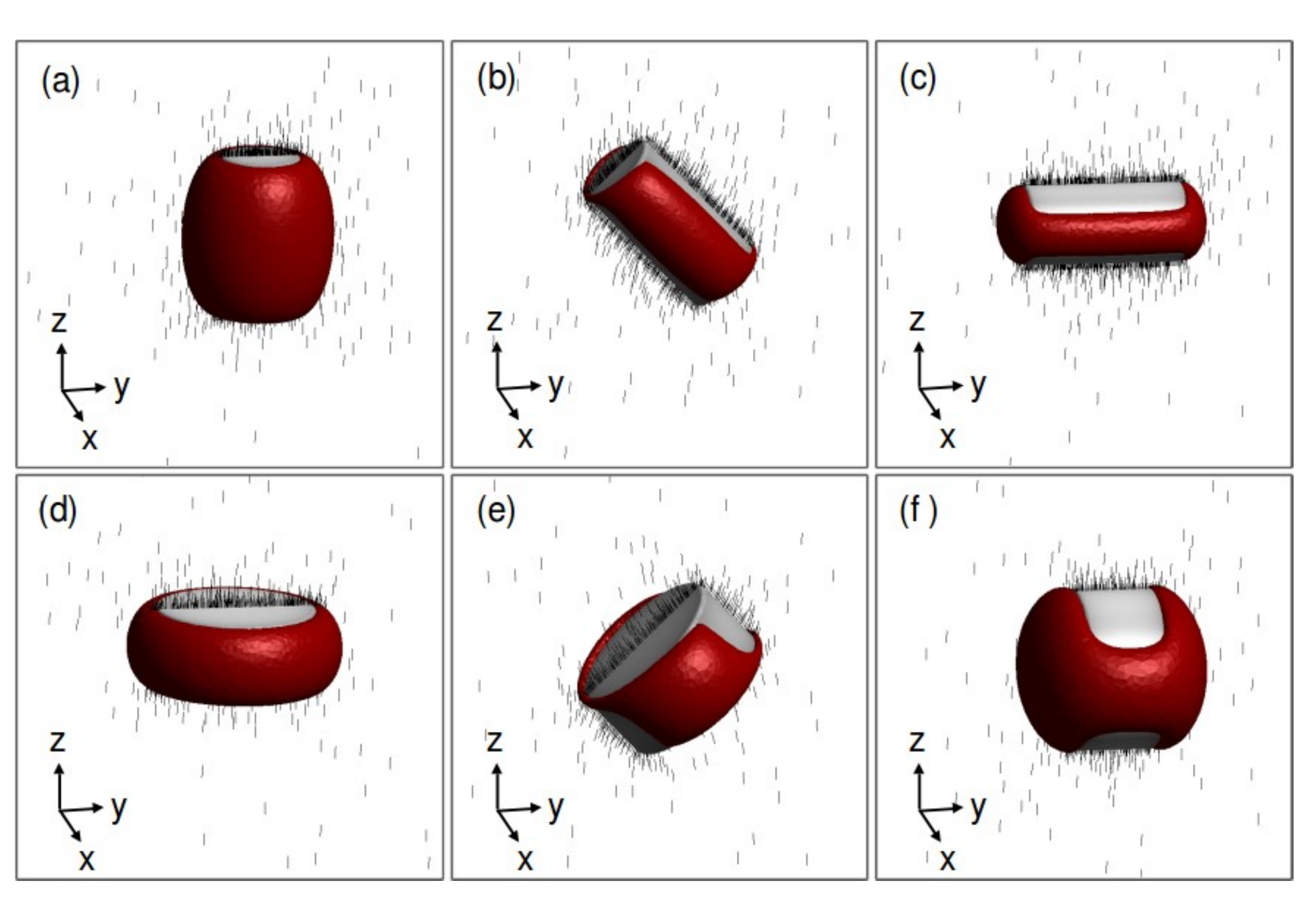}
\end{tabular}
\caption{(Color online) (a-c) Defect structures around a colloidal particle with the aspect ratio 1:2 and $R=10 \rm{nm}$ at 0, 45 and 90 degrees. (d-f) Defect structures around a colloidal particle with the aspect ratio 2:1 and $R=20 \rm{nm}$ at 0, 45 and 90 degrees. The anchoring constant is $W=5\times 10^{-3} \rm{J/m^2}$. Black dashed lines are a 2D view of the director field in the $y-z$ plane. Topological defect cores are visualized by the isosurface $S_q=0.5$. Based on the results of Fig.~\ref{result}, snapshot (c) illustrates the equilibrium defect structure related to the aspect ratio 1:2 and (d) the equilibrium defect structure related to the aspect ratio 2:1.} 
\label{nanoshape}
\end{figure*}
For the parallel alignment the defect structure is a Saturn ring that covers all the lateral surface of the cylinder, e.g. Figs.~\ref{nanoshape}(a) and (d). For the perpendicular alignment the defect structure is a ring located perpendicularly to the far field nematic that is elongated parallel to the particle symmetry axis and covers the flat ends of the particle, e.g. Figs.~\ref{nanoshape}(c) and (f). Thereby, we have only observed three defect structures for the limiting states of the nanoparticles of this geometry, e.g. Figs.~\ref{nanoshape}(a-c) or (d-f).

\begin{figure*}
\includegraphics[width=12.0cm, angle=0]{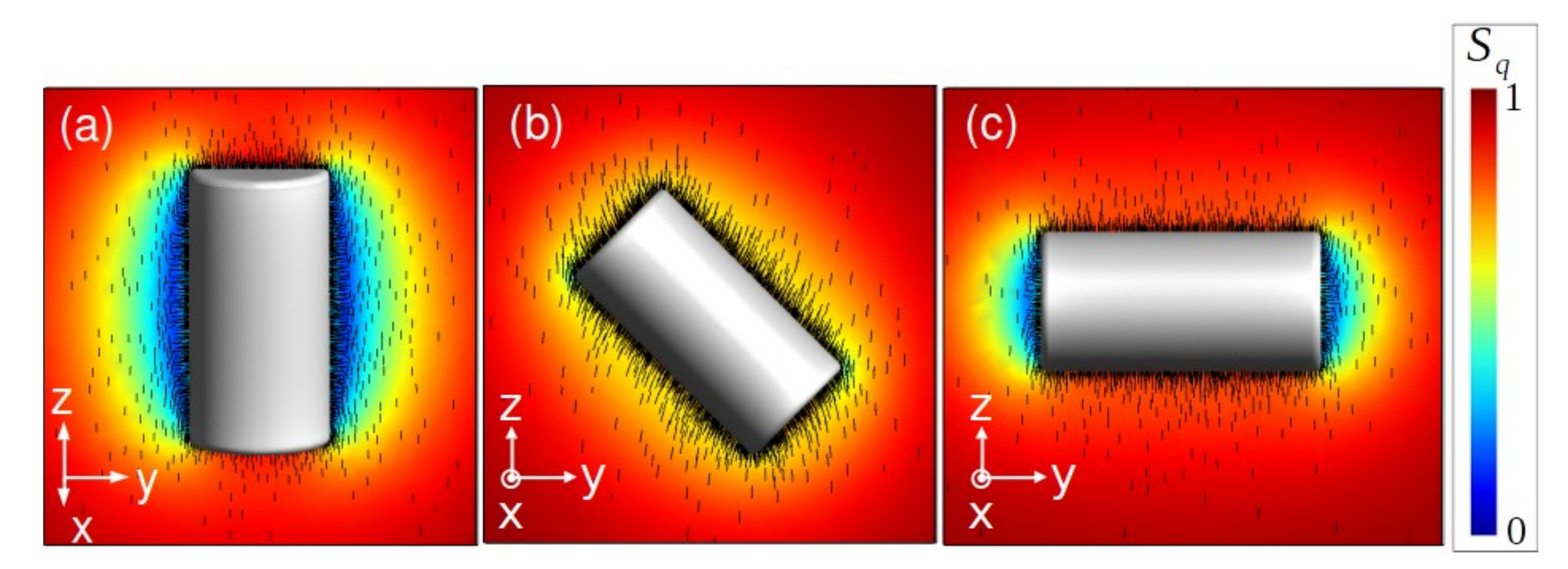}
\caption{(Color online) Director field around a nanoparticle with $R=10\rm{nm}$ and aspect ratio 1:2 for the three orientations (a) 0, (b) 45 and (c) 90 degrees and the anchoring constant $W=5\times 10^{-3} \rm{J/m^2}$. Black dashed lines are 2D view of the director field in the $y-z$ plane. The colored background represents the contour plot of the rescaled scalar order parameter in the bulk.}
\label{nanocontourplot}
\end{figure*}

To show the director field across the defect cores of Figs.~\ref{nanoshape}(a-c), as typical examples for the director field around the nanoparticles, Figs.~\ref{nanocontourplot}(a-c) include the contour plot background of the rescaled scalar order parameter in the $y-z$ plane. 

Therefore, the discrepancy between the equilibrium angles of the mesoscopic and nanoscopic colloidal particles of this geometry also manifests in their corresponding defect structures. The defect structure associated to all the asymptotic equilibrium angles of the mesoscopic colloidal particles proposed in Fig.~\ref{result} is a chair-like structure. We see that in the nanoscales the equilibrium defect structures for 0 and 90 degrees undergo a structural change. In this case, the effect of the sharp edges on the defect structures vanishes resulting in more stable structures than for the chair-like structure for oblique orientations. Therefore for the two limiting equilibrium states of the nanoparticles of Fig.~\ref{result}, it can be generally indicated that the equilibrium defect structure for a rod-like nanoparticle is typically the defect structure depicted in Fig.~\ref{nanoshape}(c) and for a discotic nanoparticle is typically the structure depicted in Fig.~\ref{nanoshape}(d).

\section{conclusion}

We studied equilibrium orientation and defect structures of a cylindrical colloidal particle with flat ends and circular cross section immersed in a uniform nematic liquid crystal. We used a finite element method for numerical minimization of the Lundau-de Gennes free energy considering homeotropic surface anchoring energy. We adopted a commonly used choice of parameters as in Ref.~\cite{FukudaPhysRevE2004}. We investigated thin discotic to long rod-like particles of this geometry with length scales ranging from mesoscales to nanoscales. We have shown that the equilibrium state is sensitive to the two geometrical parameters: aspect ratio and length scale of the particle.  

For any given aspect ratio and size of the particle there is one equilibrium angle (Fig.~\ref{equalSurface}). For a large enough mesoscopic particle, there is a specific asymptotic equilibrium angle associated to each aspect ratio, which is insensitive to further increasing the particle size. We showed that while the asymptotic equilibrium angle of a discotic particle with the aspect ratio 4:1 is $25\pm5$ degrees, upon reducing the aspect ratio, the asymptotic equilibrium angle increases and ultimately reaches $80\pm5$ for a rod-like particle with the aspect ratio 1:4 (Fig.~\ref{result}). Upon reducing the particle size to nanoscale the equilibrium angle of the particle follows a descending or ascending trend. The equilibrium angle for a particle with the aspect ratio bigger than 1:1 (discotic particles) goes to a parallel alignment, whereas the equilibrium angle for a particle with the aspect ratio 1:1 or smaller (rod-like particles) tends toward a perpendicular alignment (Fig.~\ref{result}).

Among the defect structures we have observed for a mesoscopic colloidal particle that there are two stable structures, a dipolar top-ring and a quadrupolar chair-like defect (Fig.~\ref{defectGeneral}), and that their relative stability depends on the spatial orientation of the particle (Fig.~\ref{equalSurface}). The angular intervals of stability we have obtained for these structures are in agreement with the experimental study of Ref.~\cite{TkalecMusevicRodLike2008}. In mesoscales the sharp edges of the particle play a crucial role in the relative stability of the defect structures with respect to each other. The defect structure related to the asymptotic equilibrium angles of the mesoscopic particles, presented in Fig.~\ref{result}, is a chair-like structure. We mentioned that by decreasing the particle size, the defect core size relative to the particle size increases so in the limiting states for the nanoparticles the equilibrium defect structures of 0 and 90 degrees undergo a structural change. This change in defect structures leads to a parallel quadrupolar stable defect structure for the discotic and a perpendicular quadrupolar stable defect structure for the rod-like nanoparticles. So the discrepancy between the equilibrium angle of the mesoscopic and nanoscopic colloidal particles originates from the significant differences between their defect structures.

We see to what extent the geometrical parameters of this colloidal shape can play a role in its equilibrium behavior. However, an interesting open question remains regarding the pairwise interaction between these colloidal particles for which the equilibrium configurations are also expected to be highly dependent on the aspect ratios and length scales of the particles. The significance of this study becomes clearer wherever these colloidal particles are exploited as the building blocks of a colloidal crystal in which the equilibrium orientation of a single colloidal particle may impose specific orientations to the whole layers of the crystal.
 
\section*{Appendix A}

As we are interested in studying a colloidal particle suspended in the bulk of a nematic medium far from the cell walls, one should be sure that the walls of the simulation box are far enough apart, so the walls do not interact with the colloidal particle and its neighboring defects. This helps us avoid finite size effects of the box walls on the particle's position and equilibrium angle.

To check whether we have the right size for the simulation box, we perform some numerical modeling on different azimuthal orientations and positions of the particle inside the box.
    
\begin{figure*}[!ht]
\begin{tabular}{cccc}
\includegraphics[width=14.0cm, angle=0]{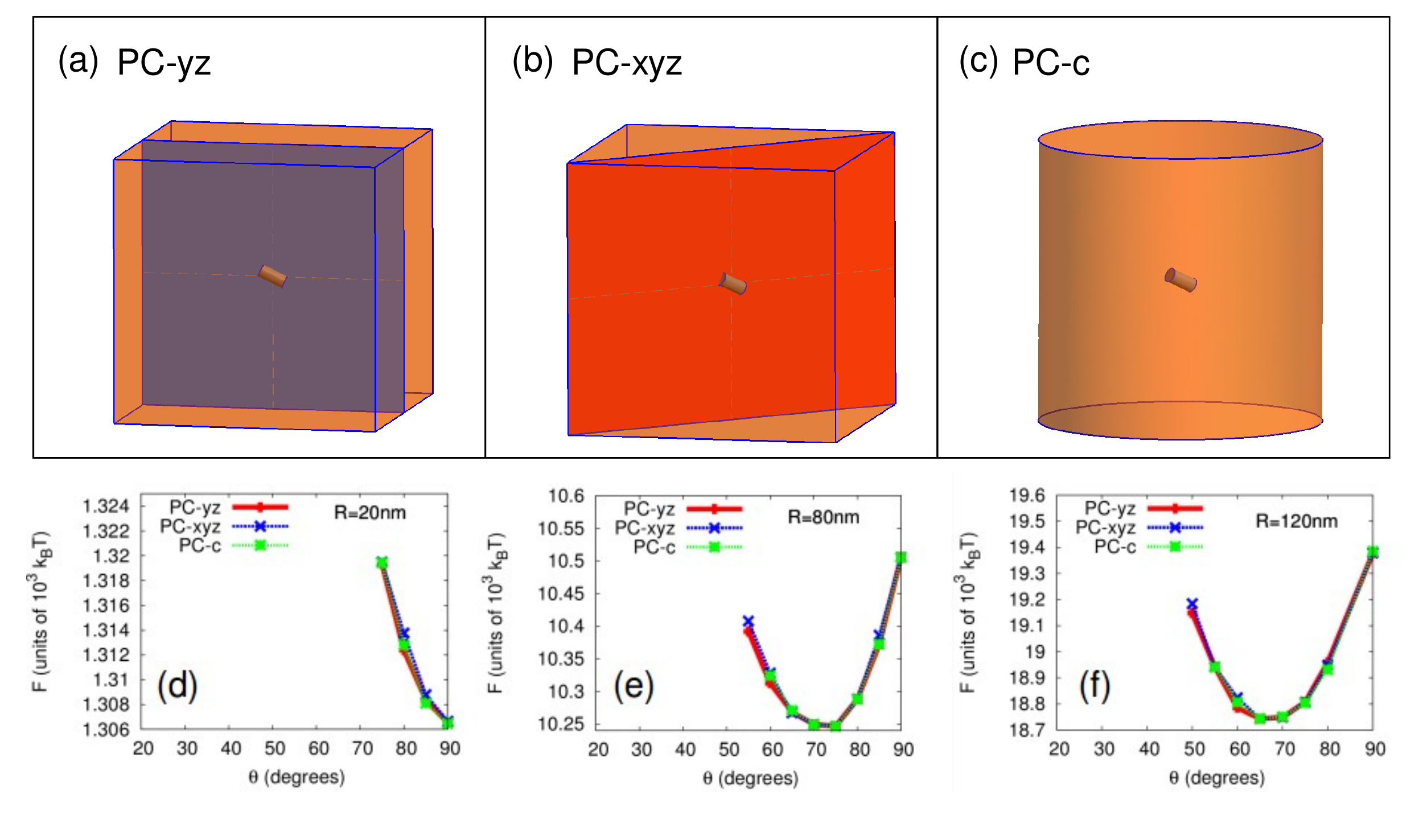}
\end{tabular}
\caption{(Color online) (a-c) Schematic representation of a colloidal particle in (a) a cubic box where the symmetry axis of the particle is in the $y-z$ plane, (b) a cubic box in which the symmetry axis of the particle is in a common plane with one of the space diagonals of the box, and (c) a cylindrical simulation box. (d-f) Free energy as a function of $\theta$ for a particle with $W=5\times 10^{-3} \rm{J/m^2}$, with the aspect ratio 1:2, and (d) $R=20 nm$, (e) $R=80 nm$ and (f) $R=120 nm$ in the three setups.} 
\label{BoxShape}
\end{figure*}

We perform modeling on a cubic (Figs.~\ref{BoxShape}(a,c)) and a cylindrical simulation box (Fig.~\ref{BoxShape}(d)). For a cubic simulation box, we consider two cases in which either the symmetry axis of the particle is in the $y-z$ plane (Fig.~\ref{BoxShape}(a)) or in a common plane with one of the space diagonals of the box (Fig.~\ref{BoxShape} (b)). Each of Figs.~\ref{BoxShape} (d), (e) and (f) illustrates the energy graphs related to a colloidal particle with a given radius and the aspect ratio 1:2 that is positioned at the center of the three above mentioned setups. One can see the energy graphs for each given particle radius in these setups overlap and their minimums are the same in an angular resolution of 5 degrees (Figs.~\ref{BoxShape} (d-f)). Therefore one can consider an azimuthal symmetry for the colloidal particle at the center of the simulation box. 

\begin{figure*}[!ht]
\begin{tabular}{cccc}
\includegraphics[width=14.0cm, angle=0]{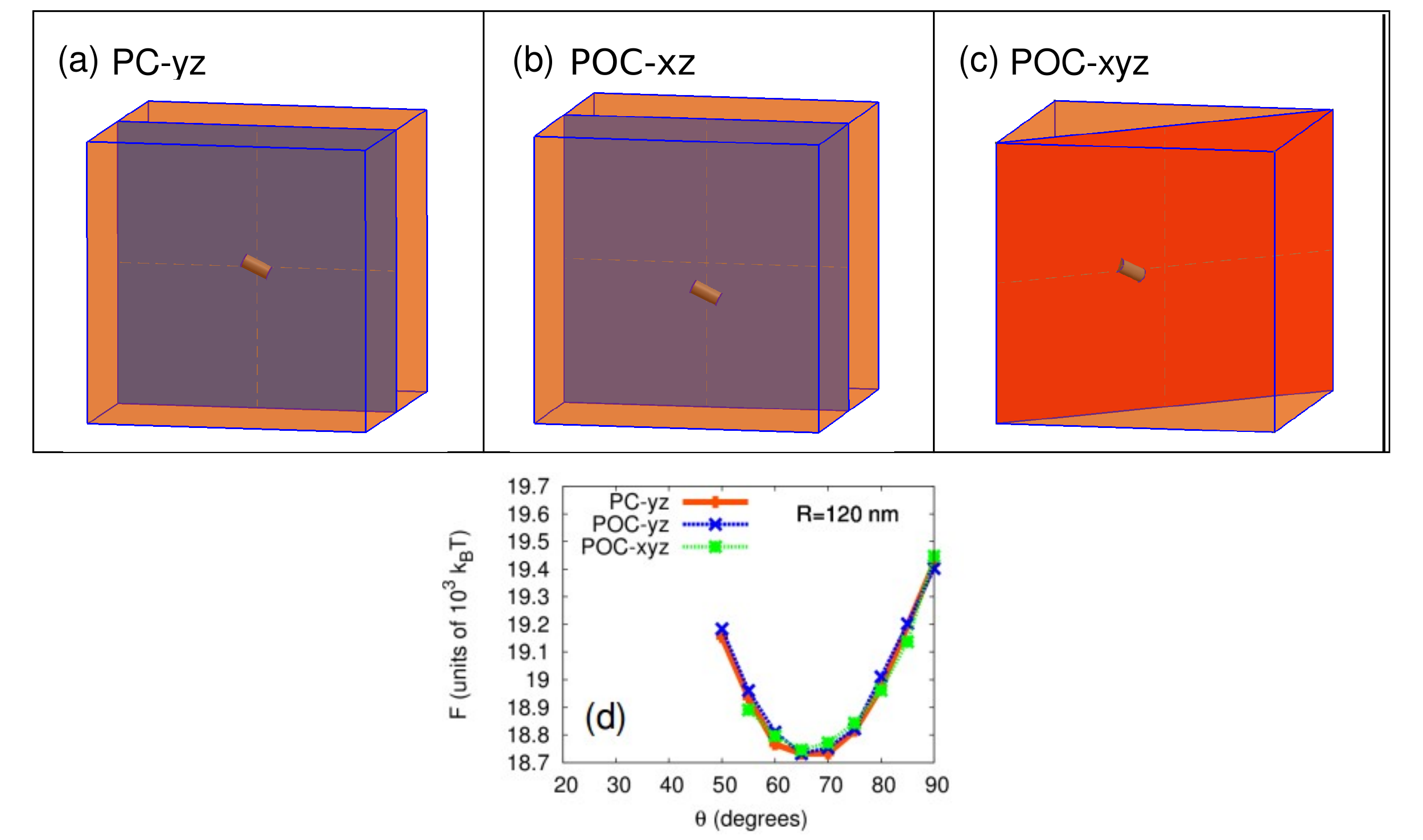}
\end{tabular}
\caption{(Color online) (a-c) Schematic representation of a colloidal particle in a cubic box. (a,b) The symmetry axis of the particle is in the $y-z$ plane and the particle is positioned (a) at the box center and (b) $500 nm$ along the $z$ axis out of the center of the box. (c) The symmetry axis of the particle is in a common plane with one of the space diagonals of the box and the particle is positioned $\sqrt{2}\ 500 nm$ along the mentioned space diagonal out of the center of the box. (d) Free energy as a function of $\theta$ for a particle with $W=5\times 10^{-3} \rm{J/m^2}$, with the aspect ratio 1:2 and $R=120 \rm{nm}$ in the three setups.} 
\label{Position}
\end{figure*}

We also perform modeling on different positions of a colloidal particle with the aspect ratio 1:2 and radius $R=120 nm$ at or near the center of a cubic simulation box. We consider the three setups Figs.~\ref{Position} (a-c) in which the particle is at (Fig.~\ref{Position} (a)) or near (Figs.~\ref{Position} (b,c)) the center of the box in different azimuthal angles. The energy graphs (\ref{Position} (d)) that are related to these setups overlap and their minimums are the same in an angular resolution of 5 degrees which shows the translational symmetry of the colloidal particle near the center of the simulation box.
Therefore, we can conclude that the simulation box is large enough so the box walls do not have observable effects on the position and equilibrium angle of the particle near the box center.

\section{Acknowledgment}
We thank Mohammad Reza Mozaffari for his valuable comments and also for development of the numerical minimization code. We also thank Ivan Smalyukh for his encouraging remarks and fruitful discussions. M.R.E. acknowledges the Center of Excellence in Complex Systems and Condensed Matter (CSCM) for partial support.

\bibliography{references} 

\end{document}